%% file: paper.tex
\documentclass[10pt,twocolumn]{article}
\usepackage{usenix,epsfig}

\usepackage{ifpdf}
\usepackage{epsf}
\usepackage{microtype}
\usepackage{graphicx}
\usepackage{url}
\usepackage{upquote}
\usepackage{color}
\usepackage{enumitem}
\usepackage{subcaption}
\usepackage{xspace}      

\usepackage[T1]{fontenc}
\usepackage[scaled]{beramono}
\usepackage{caption}
\usepackage{hyperref} 

\usepackage[utf8]{inputenc}
\usepackage{listings}
\usepackage{titling}

\graphicspath{ {./fig/} }

\newcommand{\system}{{CFlo}\xspace}

\begin{document}

\title{A Logical Approach to Cloud Federation\thanks{This paper is based upon work supported by the
US National Science Foundation through the GENI Initiative and under NSF grants OCI-1032873, CNS-0910653, and CNS-1330659.}
}
\author{Qiang Cao, Yuanjun Yao, Jeff Chase \\
Duke University
}
\date{}
\maketitle

\begin{abstract}
Federated clouds raise a variety of challenges for managing identity, resource access, naming, connectivity, and object access control.    This paper shows how to address these challenges in a comprehensive and uniform way using a data-centric approach.  The foundation of our approach is a trust logic in which participants issue authenticated statements about principals, objects, attributes, and relationships in a logic language, with reasoning based on declarative policy rules.   We show how to use the logic to implement a trust infrastructure for cloud federation that extends the model of NSF GENI, a federated IaaS testbed.   It captures shared identity management, GENI authority services, cross-site interconnection using L2 circuits, and a naming and access control system similar to AWS Identity and Access Management (IAM), but extended to a federated system without central control.
\end{abstract}

\section{Introduction}
\label{sec:intro}
\input{intro.tex}

\input{slang-apis}

\section{Overview}
\label{sec:overview}

\input{overview.tex}

\section{Logical Cloud Federation}

\input{logical.tex}


\section{Evaluation}
\label{sec:eval}
\input{evaluation.tex}

\section{Related Work}
\label{sec:related}
\input{related}

\section{Conclusion}
\label{sec:concl}

\input{concl.tex}

\bibliographystyle{abbrv}
\bibliography{silver-main,fed,bib,cloud,refs_duke,grid,trust,main,refs_genibook}

\end{document}

%% file: intro.tex
The IaaS market today is dominated by a small number of megaproviders, which compete on price and services for market position, and face disincentives to combine their offerings.  However, as the technology develops,
some speculate that cloud providers will face natural market incentives to interconnect their service offerings ({\it cloud peering}), leading to the emergence of an ``intercloud'' following the historical development of infrastructure networks including the Internet and the power grid~\cite{hughes1993networks}.  Peering enables providers to shift load to absorb demand spikes.  The IBM Reservoir project~\cite{rochwerger2009reservoir,rochwerger2011reservoir} and others popularized this model.  


An overlapping trend is the emergence of {\it multi-cloud} applications that span multiple providers.  They occur naturally in cloud peering scenarios, but
cloud adopters may also use multiple providers to manage cost or risk.  Multi-clouds are also attractive for peer-to-peer application platforms and for services that benefit from proximity to the edge of the network (cloudlet, fog, or locavore computing).  The multi-cloud model was also popularized as ``sky computing''~\cite{keahey2009sky}.  Various efforts have sought to develop stacks and standards to launch, manage and/or migrate application networks seamlessly and safely across multiple clouds: these include the Open Cloud Computing Interface (OCCI)~\cite{ogf-occi}, various research works (e.g., \cite{Williams:2012:XVO:2168836.2168849}), and Cisco's Intercloud Fabric offerings~\cite{cisco-intercloud}.  


A decade ago the research community launched major initiatives to combine network testbeds to leverage benefits of scale, diversity, geographic dispersion, and heterogeneity.  NSF GENI~\cite{Berman20145,brinn-tridentcom15} in the US and FIRE in the EU exemplify this trend.  Both initiatives have funded deployment of IaaS federations spanning many sites and providers. They also embody a third dimension of federation: they serve a common community of member researchers, requiring some form of federated identity for their users (a {\it community cloud}~\cite{nist-cloud}).  Other recent efforts take a similar approach to linking accounts across providers (e.g., \cite{chadwick2014adding}). 

These three dimensions of cloud federation---peering, multi-cloud, and community---present a common set of overlapping challenges for identity, trust, access, and governance.  Federation requires some means to
represent and certify trust relationships among users and providers, including their terms of peering.  It also places new pressure on the mechanisms to manage multi-tenancy, including
naming, ownership, and access control of protected cloud objects (machine instances, virtual storage objects, networks), and accounting and accountability for the use of resources.  The US government has identified federated/community/multi-cloud scenarios as a priority area for standards, focusing on ``credentials, namespaces, and trust infrastructure''~\cite{badger2014us-nist-cloud-roadmap}.

This paper takes a comprehensive approach to trust infrastructure for cloud federation.
We advocate a {\it data-centric} approach that captures the attributes and relationships of identities and objects, with trust and authorization based on queries over the data model.
Our approach is {\it fully decentralized}: participants exchange certificates with statements in a logic language, and issue local queries against locally cached sets of relevant assertions and declarative policy rules.  
It provides {\it end-to-end authorization}~\cite{Howell:2000}: each participant can verify for itself that its interactions comply with its policy based on statements that it has received from other parties.   We use a simplified trust logic based on Datalog---a well-studied logic language with a rigorous semantics~\cite{ceri89-datalog-dare}---within a novel system for managing certificates.


This paper uses the architecture of the GENI deployment as a model for federation.  It addresses key issues of federated identity, trust, governance, and coordination that are common to peering, multi-cloud, and community federation scenarios. 
We show how to capture the GENI trust model using logic, and extend it with access control for protected objects, using features similar to those in Amazon's Identity and Access Management (IAM~\cite{aws:iam}), but built for a multi-cloud scenario.
Finally, we show how to authorize linked private networks (virtual private clouds or VPCs) in a multi-cloud, cross-tenant peering of VPC networks, and more complex cross-federation structures.

The contributions of this paper include:

\begin{itemize}
\item {\it Specify trust and naming for federated IaaS scenarios in a way that captures the naming and trust model of the existing GENI deployment.}   We show how to use logic to frame the design issues and specify solutions in a way that is concise, precise, and verifiable.


\item {\it Demonstrate use of trust logic as an implementation technology for federated clouds.}  The logical specification is directly deployable using the SAFE framework~\cite{safe:linking16} to manage the exchange of logic content as linked certificates, and execute trust queries against assembled sets of logic statements.  (See \S\ref{sec:overview}.)

\item {\it Evaluate the performance of logical federation.}  Microbenchmarks and synthetic workloads show that key trust operations are fast enough to be practical in a deployment: they are at least an order of magnitude faster than the typical cost of the operations they protect, e.g., instantiating or linking cloud resources.

\end{itemize}

%% file: slang-apis.tex
\begin{table*}[t]
\centering
 \scriptsize
  \begin{tabular}{p{6cm} p{9cm}} \hline 
  {\bf Method } & {\bf Description}\\ \hline \hline
{\tt root.endorseAggregate(PID)} & Issue root endorsement for an aggregate (infrastructure provider). \\ \hline
{\tt root.endorseAuthority(PID, type)} & Issue root endorsement for an authority service to certify users (MA), projects (PA), or tenants ({\it slices}: SA). \\ \hline
{\tt PA.createProject(ownerPID, attributes) returns projectID}  & Create a project with owner {\tt ownerPID}, checking its permission.   This is an API call of a Project Authority (PA).\\ \hline
{\tt member(PID, projectID, role, delegatable)}  & Delegate project membership to {\tt PID} with a named {\tt role}. \\ \hline
{\tt SA.createSlice( ownerPID, projectID, attributes) returns sliceID}  & Create a slice (tenant) with owner {\tt ownerPID} in a project, checking its permission.   This is an API call of a Slice Authority (SA).\\ \hline
{\tt delegateSlice(PID, sliceID, perms, delegatable)}  & Delegate named permissions to operate on a slice. \\ \hline
{\tt Agg.createSliver( sliceID, attributes) } & Check requester's permission to instantiate virtual infrastructure at this aggregate for use by a slice. \\ \hline
{\tt Agg.sliceOperation( sliceID, type, attributes) } & Check requester's permission to perform a control action on a slice's resources at this aggregate. \\ \hline
{\tt createNameEntry(dirID, <component>, targetID)} & Create a name for {\tt targetID} in naming context of {\tt dirID}. The caller must control {\tt dirID}. \\ \hline
{\tt resolveName( <pathname>) returns ID}  & Resolve a multi-component pathname, which may cross domain boundaries. \\ \hline
{\tt createGroup() returns GID} & Create a new empty group and return its scid. \\ \hline
{\tt groupMember(GID, PID, delegatable)} & Grant membership in group {\tt GID} to principal {\tt PID}; {\tt delegatable} is a boolean that determines whether transitive delegation is permitted. \\ \hline
{\tt checkAccess( subjectID, targetID)} & Check whether a principal or group with {\tt subjectID} has the right to access an object with {\tt targetID}. \\ \hline
  \end{tabular}
  \caption{\bf \footnotesize  Simplified trust API for \system, a logical trust core for use by participants in a federated IaaS.  
 }
  \label{tab:cflo-api}

\end{table*}

%% file: overview.tex
This paper describes a trust core for cloud federation using logic (``\system'') based on the SAFE logical trust framework.  SAFE factors trust concerns out of the cloud services and tools, and isolates them in application-supplied {\it logic scripts}.  We implemented \system in about 600 lines of SAFE scripting code, including logic templates for all credential formats, exemplary policy rules, and compliance queries.  The scripts implement a {\it trust API} to manage credentials and make trust decisions (see Table~\ref{tab:cflo-api}).    The scripts run in a SAFE interpreter engine that is local to each participant and under its direct control. 

This paper is not about SAFE itself, which is the topic of a companion paper~\cite{safe:linking16}.  Rather, it is about using logical trust to address a range of issues in cloud federation.   SAFE provides an exemplary trust logic language and system that enables us to evaluate how \system would perform in a real deployment. 

{\bf GENI.}  The design of SAFE was motivated by our experience in applying logical trust in the development of GENI.    Although GENI was conceived as a network testbed, it is best understood as a federation of autonomous IaaS providers (``aggregates'') linked by various trust relationships and agreements.   GENI serves a community of registered researchers with various institutional and project affiliations.  Each provider has various policies governing client access.  These policies consider endorsements and delegations of trust among the participants, including a root trust anchor that certifies the aggregates and various {\it authority} services to govern membership and coordination.   In this respect GENI is representative of federated cloud systems in general, although there are differences in terminology.

GENI uses the {\it slice} abstraction for multi-cloud scenarios, first introduced in PlanetLab~\cite{planetlab:osdi06}.   A slice is a logical container for a set of virtual resources (e.g., VMs, network links) that may span multiple providers and are allocated and used for a common purpose.  A {\it sliver} is a typed virtual resource unit that is provisioned from a single aggregate and is named and managed independently of other slivers.  Each sliver is bound to exactly one slice at the time that the sliver is created.  
Users may link slivers from multiple providers to form end-to-end environments (slices) for networked applications. 


{\bf ExoGENI.}  One goal of \system is to extend ExoGENI~\cite{baldine12:exogeni,baldinexog2016} to enable richer forms of cross-tenant interaction, including discretionary access control for slivers and cross-slice network peering.   ExoGENI is a federation of xCAT/OpenStack cloud clusters on 20 campuses, linked by the Internet2 AL2S and ESnet network circuit fabrics.  It supports elastic multi-cloud slices with private networks (VPCs constructed by {\it stitching} VLANs at layer2) that may be tenant-managed via OpenFlow.  It automates end-to-end assembly of the slice VPC dataplane across multiple providers~\cite{globecom10-ndl}.  To do this, it provisions cross-cloud circuits on demand, bridging among circuit fabrics at exchange points (e.g., at Starlight) as needed. 
As of February 2017 ExoGENI has supported over 56,000 experiments/slices submitted by more than 1400 distinct users.


To integrate \system into ExoGENI, we must modify its control servers to invoke \system APIs in a local SAFE engine to check each action for compliance with a trust policy before executing it.   Beyond enabling new functionality, \system can place the existing security and peering mechanisms in ExoGENI on a more uniform and extensible foundation. 




{\bf Scope.}   The \system trust scripts implement the GENI trust core in logic.  For this paper, we added script support for user groups, hierarchical names, and access control for cloud objects (e.g., ACLs for virtual network links), all modeled on AWS Identity and Access Management (AWS-IAM~\cite{aws:iam}).  We also added logic for authorization that takes place during VPC stitching in ExoGENI, and combined it with ACLs to enable network peering among tenant VPCs by mutual consent.   This paper does not address how the underlying operations (VPCs, L2 stitching) are implemented and orchestrated; refer to~\cite{baldinexog2016,chaseretr2016}.  We also do not address resource discovery, resource brokering, or payment models.   ExoGENI is based on our earlier work on these topics (e.g., \cite{fu03}).  Integrating these mechanisms with logical trust is future work. 



\subsection{Building with Trust Logic}

Logical trust has a long history in the research community~\cite{abadi93:accesscalculus,lampson92:authentication,spki-rfc2693,jim:sd3:2001,Li02:RT,detreville:binder:2002,Sendlog:Abadi:2007,abadi08, Schneider:2011:NAL:1952982.1952990}.  (See \S\ref{sec:related}.)  Like many logical trust systems, SAFE is a credentials-based 
PKI system.  Each principal has a keypair to authenticate its requests and sign any credentials that it issues.   Participants exchange security assertions and policy rules as semantically rich logic-based certificates, and run a local engine to
generate proofs of policy compliance end-to-end.   Certificates have a period of validity that is checked along with the signature on import or use: the prover sees only logic content that is fresh and authentic.


SAFE's trust logic is based on Datalog~\cite{ceri89-datalog-dare}, a rigorously defined and extensively studied general-purpose logic language that is a subset of Prolog, a popular language for logic programming with a standard syntax.  It adds a modal operator {\bf says} to Datalog, enabling its direct use as a logic of belief and attribution, following Binder~\cite{detreville:binder:2002}, SD3~\cite{jim:sd3:2001}, and SENDLOG~\cite{Sendlog:Abadi:2007}.  

Datalog content consists of atomic statements (atoms)  and rules built up from atoms and the logical operators
conjunction and implication.  An atom is a predicate symbol applied to a list
of parameters, which may be variables or term constants representing principals,
objects, or values.  Predicate symbols are user-defined: they may represent properties, attributes, roles, relationships, rights,
powers, or permissions.  Atoms whose parameters are term constants (ground) represent simple assertions equivalent to a row of a database table.  Rules embody implication and may contain variables.  A rule has a head and a body.  The head of a rule is a single atom.  The body is a sequence of atoms ({\it goals}) separated by commas, which indicate conjunction: all of the atoms in the body
must be true for the rule to ``fire''.    A rule allows the prover to infer that the head is true
for some substitution of its variables with constants, if the body is true
under that substitution.




In Datalog-with-says, every atom has a first (prefix) parameter representing a principal
who {\bf says} it (the {\it speaker}).   If the parameter is omitted, it defaults to the current principal ({\tt \$Self}).
In this way, a statement about a principal  naturally represents a delegation or endorsement that is restricted by the speaker and predicate; another principal considers the statement only if it has a policy rule with a matching goal, conferring trust in the speaker.

Datalog-with-says is sufficiently powerful to represent common access control features hierarchical naming, nested groups, roles and other attribute assertions, ACLs, and capabilities.   Delegations may be constrained by a predicate/role and by parameters (e.g., {\it ``Alice owns file F''}).  Conjunctive policy rules permit reasoning from multiple attributes of a principal or object, and policies are mobile: they may be passed in certificates.

SAFE defines conventions for self-certifying term constants (IDs) to name principals and objects.
A principalID is a SHA hash of the principal's public key, following SPKI/SDSI~\cite{spki-rfc2693}.  
All statements in a valid certificate must have a speaker ID that matches the issuer who signed the certificate.
Each object named in a logic statement has some principal who is its {\it controlling authority}.  The objectID consists of an identifier (a UUID/GUID) chosen by its authority, concatenated with the authority's principalID to form a {\it self-certifying identifier (scid)}.
SAFE scripts use a builtin function {\tt rootID} to obtain a scid's controlling principalID.   Self-certifying IDs ensure that parties have distinct names for their objects, and a malicious principal cannot ``hijack'' another's names.
In this way logical trust extends conventional identity-based PKI security
to incorporate rich statements about principals, objects and
their security attributes, and avoids the need for a global naming root.

\subsection{SAFE Logic Scripting}

SAFE synthesizes elements from previous trust logic systems and extends them with additional system support to enable practical deployment.  The novel elements of SAFE include a scripting language to insulate applications from logic concerns, and an interface to a shared key-value store (e.g., a DHT), which stores authenticated logic content as signed certificates in a native SAFE format.  Certificates in the store are indexed by self-certifying links ({\it tokens}), and can be written only by their issuers.  The application trust scripts contain parameterized logic templates to generate certificates easily, and also to link certificates to construct DAGs programmatically as a side effect of delegations.

This use of {\it certificate linking} simplifies discovery and retrieval of the content relevant to a trust decision.  The certificate links (tokens) also enable pass-by-reference and caching of certificate content at the authorizers.  The shared certificate store enables an issuer to update or revoke its certificates by their tokens, addressing common PKI concerns.

Scripting is organized around the abstraction of {\it logic sets} --- sets of logic statements that represent credentials, delegations, endorsements, and policies.   Scripts use templated constructors ({\tt defcon}) to construct and modify sets and link them to form unions.

A principal may issue ({\it post}) its logic sets and share them by reference; posted sets are materialized as certificates spoken by the issuer and signed under its keypair.  
A posted set is accessible to any client that knows its token, but only its issuer can modify it.  Scripts name their locally constructed sets with arbitrary string names ({\it labels}); the token is a SHA hash of the issuer ID and the label.   Thus tokens are ``unguessable'', but anyone who knows the label and the issuer's public key can synthesize a set's token.   Some \system script actions (e.g., name resolution) obtain links in this way.

SAFE {\it guard} scripts ({\tt defguard}) combine linked sets to construct query contexts, and issue queries to check policy compliance for trust decisions.    SAFE fetches a certificate when a guard references a logic set by its token.   After validation SAFE extracts the semantic content of the certificate into a logic set cached in an in-memory {\it set cache}.   The scripts deal only with the semantic content: the SAFE runtime encodes and decodes logic material, handles cryptographic operations, and performs fetch, retrieval, and caching automatically and transparently.  

We assume that all \system participants run scripts with common logic/certificate templates, although they may install different policy rule sets.  (This assumption assures interoperability, but it is not required for security.)  Each participant's TCB includes the interpreter and scripts, which are all under its direct control: authorization is naturally end-to-end~\cite{Howell:2000}.

%% file: logical.tex
\ifpdf

\begin{figure*}[t!]
\begin{center}
\epsfig{file=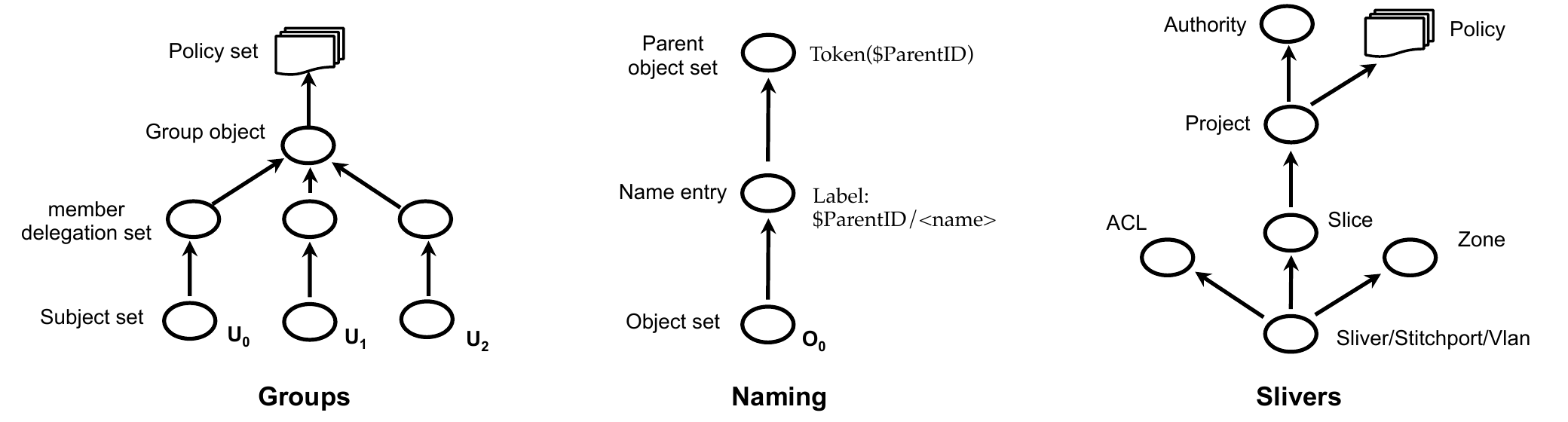, height=3.4cm}
\caption{\bf \footnotesize 
Groups, naming, and slivers in \system and their set linking patterns. In \system groups, links traverse group member delegation and group policies. In \system naming, links traverse the name entry delegation and parent objects.  Slivers link to ACL set, zone, and the containing slice. A slice also links to its containing project, which further links to a policy package and to credentials of its endorsing authority.  The logic scripts form the graphs naturally by leaving ``back links'' as a side effect of delegation.
}
\label{fig:CFLO-name-groups}
\end{center}
\end{figure*}

\begin{figure}[t!]
\begin{center}
\epsfig{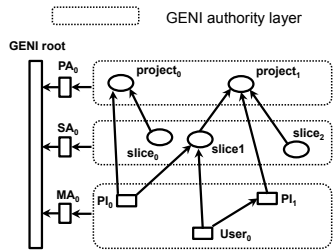}
\caption{\bf \footnotesize 
Linking patterns for \system sets/certificates in the GENI-derived trust model.  Users, projects, and slices link to their endorsing authorities; slices link to their projects; users and leaders (PIs in the figure) link to the projects they own and/or have membership in.
}
\label{fig:geni-fed-linking}
\end{center}
\end{figure}

\fi


Cloud peering and multi-cloud models raise the question of how providers are qualified to serve users, and the degree of trust that users have in them.
The community model raises the question of how
providers authenticate consumers (users), qualify them for service,
and hold them accountable for their actions in the cloud.
A user's privilege at a provider is based on membership and roles within organizations, 
relationships and agreements among organizations and providers,
and community policies and provider policies for authorization and resource management. 
These affiliations, roles, relationships, and policies may be dynamic.

This trust information flows from organizational processes outside the scope of the trust system, but the system must capture it and reason about it.  Key aspects of trust in federated systems reduce to choices about whose assertions to believe or whose commands to accept.
Trust logic offers a formalism to represent these choices.  This section presents examples from \system to illustrate this power and flexibility and to expose key issues and techniques for cloud federation.   They also illustrate the role of linking to organize sets and certificates in SAFE; some links and labels are omitted or simplified for brevity.
Figure~\ref{fig:CFLO-name-groups} and~\ref{fig:geni-fed-linking} illustrate some linking patterns relevant to this section.




\newcommand\Small{\fontsize{6.6}{7.2}\selectfont}
\newcommand\Smaller{\fontsize{6.2}{6.8}\selectfont}
\newcommand*\LSTfont{\Small\ttfamily\lsstyle}
\lstset{basicstyle=\LSTfont,breaklines=true,xleftmargin=0.4cm}

\begin{lstlisting}[language=Prolog,caption=\footnotesize Example of a logic template in a constructor: endorse a principal ({\tt ?User}) as a registered member of the federation and as a leader who is empowered to create and manage project groups.,label=code:fedUser]
defcon endorseLeader(?User) :- {
  fedUser($User).
  fedLeader($User).
}.
\end{lstlisting}

Listing~\ref{code:fedUser} shows how to generate a logic set from a template in a script.  This rule defines a set constructor ({\tt defcon}), which returns a logic set formed by substituting script variables in a template.
Each item listed within the brackets is a logic statement with an application-defined predicate
asserting an attribute for this user: the value
of the {\tt ?User} variable resolves to the user's PrincipalID and is substituted in the template using the \$ operator.

The set is materialized as a certificate signed by its issuer, the principal executing the script ({\tt \$Self}).  If another principal (an authorizer) imports the certificate, its prover sees each statement within the set as spoken by the authenticated issuer.  Any principal may issue an endorsement, but an authorizer considers them
according to its policy rules to determine whether or not to accept any given statement based on the identity of its speaker.

\begin{lstlisting}[language=Prolog,caption=\footnotesize Policy rules
  to accept user endorsements from any principal that is endorsed as
  Member Authority by a federation root trust anchor accepted by the
  local configuration.,label=code:memberAuthority]
defcon registeredUserPolicy() :- {
  fedUser(?U) :- mAuthority(?MA), ?MA: fedUser(?U).
  fedLeader(?U) :- mAuthority(?MA), ?MA: fedLeader(?U).
  mAuthority(?MA) :- fedRoot(?R), ?R: mAuthority(?MA).
}.
\end{lstlisting}

In this case, policy rules reject such endorsements unless they are issued by a Member Authority service endorsed by a federation trust anchor, as specified by the policy in Listing~\ref{code:memberAuthority}.
These statements are policy rules: the terms {\tt ?U}, {\tt ?MA}, and {\tt ?R} are variables.  These rules specify conditions to accept that a given principal is a registered user in the federation with the attribute {\tt fedUser} and/or {\tt fedLeader}.   The first rule concludes that a {\tt ?U} (whose value is the principalID of a user) is a member only if some principal {\tt ?MA} {\bf says} that {\tt ?U} is a member {\bf and} {\tt ?MA} is locally accepted as a Member Authority (MA).

The other rules in Listing~\ref{code:memberAuthority} have a similar structure.  These rules are examples of {\it attribute-based delegation}: they accept statements based on the attributes of their speakers.  The third rule says that a principal is accepted as an MA only if a {\tt fedRoot} trust anchor says that it is an MA.   A server's operator may configure its accepted trust anchors by asserting them as {\tt fedRoot} facts. 
Given a certificate from a configured root anchor endorsing an MA, and a certificate from the MA endorsing a user, these rules accept the user endorsement.

This example shows how to establish authority services in a federation to certify users (and providers) and attest to their attributes.   The MA bases its assertions on external information about the users, e.g., from a Web identity (SSO) protocol such as OAUTH or Shibboleth/SAML~\cite{shib04}.  For example, GENI runs a portal service that harvests attributes about each academic user from a Shibboleth identity provider (IdP) at the user's institution.   Once logged in, the user may supply a profile and accept required conditions.    If the user provides its key hash, the portal may issue endorsements to approve its principalID as a federation user ({\tt fedUser}) or a research team leader  ({\tt fedLeader}) based on attributes supplied by the IdP (e.g., user is a faculty member).

In this example the participants accept authorities endorsed by a common trust anchor, but they might instead configure local policies for accepting authorities.  They might select a locally accepted set, or subscribe to multiple root anchors.

\begin{lstlisting}[language=Prolog,caption=\footnotesize Set/certificate constructor for a typed user-defined group object owned by a specified subject\, and linked to a set of policy rules controlling delegation of rights to this object.,label=code:creategroup]
defcon createGroup(?SubjId, ?GroupId, ?Policy) :- {
    owner($SubjId, $GroupId).
    group($GroupId). 
    link($Policy).
}.
\end{lstlisting}

\begin{lstlisting}[language=Prolog,caption=\footnotesize
  Set/certificate constructor to delegate membership in a group to a
  subject.  A boolean indicates whether the receiver may delegate it further.,label=code:groupmember]
defcon addMember(?GroupId, ?SubjId, ?Delegable) :- {
    groupMember($GroupId, $SubjId, $Delegable).
    link($GroupSetRef).
}.
\end{lstlisting}

\subsection{Groups, Names, and Authority}

It is often useful for participants to assert their own attributes about one another.  For example, AWS-IAM provides a rich API for user-defined groups, a common basis for access control.  In \system, any principal may declare a group as an object, and issue certificates granting ownership or membership in the group with named privileges or roles.   Members may delegate their rights to others transitively using a capability model.  \system uses a standard set of logic rules (not shown) to govern this delegation by checking endorsement chains similar to the rules in Listing~\ref{code:memberAuthority}.  The rule set is linked to the group, and may be customized, e.g., to manage specific roles or privileges.  

Listing~\ref{code:creategroup} shows a constructor to create a group, and Listing~\ref{code:groupmember} shows a simple constructor to grant membership in a group.  When invoked with concrete IDs as parameters, these constructors return sets with logical assertions declaring the existence of the group, its owner and members, and its governing policy set.  These sets may be posted as linked certificates, enabling other parties to query group memberships, e.g., to control access.

\begin{lstlisting}[language=Prolog,caption=\footnotesize Constructor to install a human-readable name for an object.  The name is resolvable relative to the parent object\, which acts as a directory.,label=code:createname]
defcon createName(?Name, ?ObjectID, ?ParentID) :- {
    link(token($ParentID)).
    nameEntry($Name, $ObjectID, $ParentID).
    label("$ParentID/$Name").
}.
\end{lstlisting}

A principal may also issue a symbolic string name for any ID, specifying any object that it controls to serve as a parent context for the name (i.e., a directory).   Listing~\ref{code:createname} shows a constructor that generates a set with the name entry as a logic statement.   When posted as a certificate, its token is hashed from the parentID and name string.   A resolution procedure can synthesize this token and retrieve the set to look up the objectID by its name, given the parentID.  The named object may itself act as a parent/directory for another component of a hierarchical pathname.

These primitives enable a common namespace of groups and other objects that span principals (naming domains).   Because the named object in Listing~\ref{code:createname} may be controlled by a different principal, the name space is federated in a structure equivalent to DNSSEC~\cite{jim:sd3:2001}.  \system includes a script to resolve and certify hierarchical names relative to a root object chosen by the caller.  To share a common name space, participants must choose a common root by some convention, as with DNSSEC.  A federation may certify the naming authority as with the Member Authority example in Listing~\ref{code:memberAuthority}.  A common naming root enables \system to create a name space equivalent to the URN conventions used in GENI~\cite{brinn-tridentcom15}, which relies on an external service---DNS.

In the same way, a federation may wish to designate authorities to control the creation of groups used for federation-mandated access policies, as opposed to user-defined policies.  GENI takes this approach to manage an authoritative space of {\it project} groups to organize user activity.  Creating a GENI-sanctioned project is an action reserved to a designated authority role---Project Authority (PA)---which serves requests to create projects and is endorsed by the federation root.

A PA restricts the creation of projects to qualified users---for example, users qualified as research team leaders as shown in Listing~\ref{code:fedUser} by the {\tt fedLeader} attribute.  This is important because all user activity in GENI is associated with a project, and the project leader is accountable for that activity.

\begin{lstlisting}[language=Prolog,caption=\footnotesize Policy guard
  used by a federation-approved Project Authority (PA) to authorize a request to create a project group.,label=code:projectpolicy]
defcon projectPolicySet() :- {
    approveProject(?Owner) :- fedLeader(?Owner).
    label("policy-name").
}.

defguard createProject() :- {
    link($AnchorSet).
    link(token("policy-name")).
    link($BearerRef).
    approveProject($Subject)?
}.
\end{lstlisting}

Listing~\ref{code:projectpolicy} gives a simple example of a policy guard to enforce this restriction, including a simple policy rule set linked by a standard label.  The PA server invokes the {\tt defguard} action in Listing~\ref{code:projectpolicy} on a request.  The guard creates a set of statements, similarly to the constructor examples, and then issues an {\tt approveProject} query against this set (the query context).  The query is a guard condition: if it is provable, then the request is approved, else it is denied.   The policy rule at line 2 concludes {\tt approveProject} if the project owner (in this case the {\tt \$Subject} who issued the request) is a {\tt fedLeader}, as governed by the rules in Listing~\ref{code:memberAuthority}.

The guard imports all of the needed rules through links to the policy set (line 8) and to the authorizer's {\tt AnchorSet} (line 7), its set of configured facts (e.g., trust anchors) and rules, including the rules in Listing~\ref{code:memberAuthority}.  It also imports a standard {\tt BearerRef} variable, which resolves to a token that must be passed by the requester.   For example, if the requester passes a link to the certificate issued in Listing~\ref{code:fedUser} (or any set that links to it), then the guard fetches the MA's endorsement assertions into the context.   These in turn link to the root's certification of the MA, enabling the query to succeed.  This example illustrates the power of certificate linking in assembling a query context for a guard.  Figures~\ref{fig:CFLO-name-groups} and~\ref{fig:geni-fed-linking} illustrate linking patterns for \system.

\subsection{Resource Access}

To determine whether or not to approve a given request for resources, a provider policy may consider the purpose and authority of the request as well as the identity and attributes of the requester.   In GENI, every request for resources (i.e., a sliver) is linked to a slice, and every slice is linked to a project.  The project and slice are objects that may have arbitrary attributes associated with them (e.g., high priority, top secret) by their controlling authorities.

Since a provider's policies may use these attributes to govern resource access and accounting, the provider must accept the authorities that certify them, e.g., they must be federation-approved like the MA in Listing~\ref{code:memberAuthority} and the PA in Listing~\ref{code:projectpolicy}.  GENI defines a third authority role (Slice Authority, SA) to approve creation of slices.  As with all of the authority types, there may be many SAs in the federation.  Providers may choose which authorities to accept, and they may consider attributes of the authorities as well in their policy decisions.

\begin{lstlisting}[language=Prolog,caption=\footnotesize Guard rule
  for a federation-approved Slice Authority (SA) to authorize a request to create a slice.,label=code:sliceguard]
defcon guardPolicySliceCreate() :- {
    approveSlice(?Subj, ?Proj) :-
    ?PA := rootID(?Proj),
    projectAuthority(?PA),
    ?PA: group(?Proj),
    ?PA: memberPriv(?Subj, ?Proj, instantiate, _).
}.
\end{lstlisting}

Listing~\ref{code:sliceguard} shows an exemplary policy rule used by an SA guard to create a slice.
To approve a request, the SA must be convinced that it is associated with a valid project group (line 5) approved by an eligible PA (line 4), and that the subject has permission within the project to bind a slice to it according to the project policies (line 6).

Listing~\ref{code:sliceguard} illustrates the use of the {\tt RootID} builtin to obtain the ID of the controlling principal (the PA) for the project from its object ID.  Like all objects, the project ID is named by a self-certifying identifier that incorporates the principal ID.  It also illustrates how a policy rule can delegate policy control to rules issued by another principal (the PA) and evaluated locally ({\it policy mobility}).  The PA's policy rules for the project group are spoken by the PA and linked from the project set (Listing~\ref{code:projectpolicy}, line 8); the guard fetches them when it pulls the closure of the requester's {\tt BearerRef}: the  {\tt BearerRef} links to its membership certificate (Listing~\ref{code:groupmember}), which links to the policy set at line 4.  Trust logic enables an authorizer (the SA) to evaluate policy rules spoken by another party (the PA) to determine if that party ``says or believes'' that the request is valid according to its own policies (lines 5-6 of Listing~\ref{code:sliceguard}); attribution is sound across inference.  Of course, the authorizer may add restrictions of its own.

\begin{lstlisting}[language=Prolog,caption=\footnotesize Guard policy for a cloud provider to authorize control of a slice.  The caller must be a member of the slice with suitable privilege by the policy of its controlling authority (SA).,label=code:sliceControl]
defcon guardPolicySliceControl() :- {
    approveSliceControl(?Subject, ?Slice) :-
        ?SA := rootID(?Slice),
        ?SA: slice(?Slice, ?Project),
        sliceAuthority(?SA), 
        ?SA: memberPriv(?Subj, ?Slice, control, _).
}.
\end{lstlisting}

Similarly, Listing~\ref{code:sliceControl} shows an exemplary policy used by a provider as a condition to authorize a caller to control a slice, e.g., to approve a resource request for the slice.  The slice is also associated with its own group whose members have various roles in the slice, and may obtain these privileges through group delegations according to the policy of the controlling principal---the slice's SA.  As with all groups, the members may have been endorsed by different MAs in this federated system, e.g., they may be associated with different institutions. 

Once a request for cloud resources is authorized, a provider may limit, delay, or reject the request based on a separate resource allocation policy.  This policy may consider arbitrary attributes of the user identity, slice, or project, and/or attributes of their approving authorities.   For example, a simple policy might be to treat projects as a unit of accounting, analogous to {\it accounts} in AWS.

\subsection{Protected Objects and ACLs}

Cloud services enable their users to assemble sets of virtual resources, including VMs, images, storage buckets, and network links---slivers.  Advanced cloud systems like AWS enable account owners to control access to these resources for users within their accounts.  (AWS accounts are identity domains and also are similar to projects in that all slivers are linked to an account.)  AWS-IAM allows account owners to organize their objects within a hierarchical name space, manage groups of users, and attach policies to groups of users and objects (e.g., objects with common name prefixes) governing access on the basis of user and group identities.

While AWS is controlled by a single provider, the group and naming mechanisms outlined above are sufficiently powerful to extend these features to a federated system.   \system enables users to manage their own groups and control access to their objects on the basis of those groups or groups created by others.  The objects and groups may originate anywhere within the system.  What is needed is to add fine-grained ACLs to objects, including slivers.

\begin{lstlisting}[language=Prolog,caption=\footnotesize Guard policy for a cloud provider to authorize control of a protected cloud object (a sliver of a slice).  The caller must have control privilege over the containing slice.,label=code:slivercontrol]
defcon guardPolicySliverControl() :- {
    approveSliverControl(?Subj, ?Sliver) :-
        sliverOf(?Sliver, ?Slice),
        ?SA := rootID(?Slice),
        ?SA: memberPriv(?Subj, ?Slice, control, _).
}.
\end{lstlisting}

GENI bases access to a sliver on a requester's role in the containing slice.  
Listing~\ref{code:slivercontrol} gives an example of a guard rule for control of a sliver under the GENI model.  It simply checks that the requester has control privilege in the sliver's slice under the policy of the controlling SA.   The structure is similar to Listing~\ref{code:sliceControl}, but it illustrates the association of the sliver with its slice (line 3).  \system asserts a {\tt sliverOf} statement in a set when a sliver is created (see below), along with a name and other attributes.   The sliver set links to a name entry, an ACL set, and the containing slice; the closure of all of these sets are fetched into the context for guard operations involving the sliver.

\begin{lstlisting}[language=Prolog,caption=\footnotesize Guard policy for a provider to authorize access to a sliver.  If it is a network sliver from another provider then that provider must be locally accepted as a peer aggregate.  The sliver provider must grant access according to its policy.,label=code:sliveraccess]
defcon guardPolicySliverAccess() :- {
    approveSliverAccess(?Subject, ?Sliver) :-
       ?CP := rootID(?Sliver),
       aggregate(?CP),
       ?CP: sliverPriv(?Subj, ?Sliver).
}.
\end{lstlisting}

An ACL is a logic set containing a list of policy rules each stating that a specified identity or group (or a conjunction/intersection of groups) has access to the protected object.  Listing~\ref{code:sliveraccess} 
shows how \system checks access to a sliver according to its ACL, by querying an access condition ({\tt sliverPriv} at line 5).  Note that the sliver may be associated with a different provider: the rule identifies the provider (line 3), validates it as a qualifying peer (e.g., endorsed by a common root anchor), and checks {\tt sliverPriv} access according to its policy.  The ACL is a set of rules to infer {\tt sliverPriv}.  
This access is based on any rules installed in the ACL set, or control over the containing slice.
Rules are added to the ACL by a guarded operation that requires control over the sliver. 

\subsection{Stitched Interconnection}

While it may seem odd to operate on slivers across provider boundaries in Listing~\ref{code:sliveraccess}, \system uses this to authorize stitching operations on cross-aggregate network links---dynamic circuits.   ExoGENI defines a sliver type called {\it stitchport} to represent a logical network endpoint that is stitchable at an adjacent switch.  Abstractly, a stitchport occupies some tag that is unique among other endpoints in a network zone of location that is controlled by a single provider.  An endpoint of a network link---a locally attached circuit or a slice dataplane network (VPC)---is assigned a VLAN tag that is unique within the containing provider's network, which may be zoned for scaling.  VLANs may be stitched to node slivers (VMs), or to other VLANs (with tag translation).  Cross-provider stitching occurs at zone borders.  {\it Example code and explanation omitted for space: one column.}

A key element of this scenario is that adjacency implies trust among the adjacent providers, who are cooperating to establish a virtual network spanning providers within a federation.  As with examples above, the provider who executes the operation is trusted to respect and enforce the federation policy.  In general, providers control their own domains, and 
any compliance with external policies is inherently voluntary.  Participants respect these policies because they agreed to do so as a condition of their cooperation.  The federation trust structure and root endorsements ensure that they do not expose themselves to other parties who are not trusted to respect rules within the federation.

%% file: evaluation.tex
We evaluate logical federation by running representative workloads on a cluster of SAFE instances loaded with \system  trust scripts for cloud federation.   Each SAFE instance is a Scala process serving a REST API to invoke its trust scripts.   For these experiments, we evaluated the cost of logical trust with a multi-threaded load generator process that invokes the \system trust scripts directly according to synthetic request mixes designed to demonstrate and stress specific functions and behaviors in a federated cloud.   The SAFE engine and scripts handle all certificate generation, validation, and logical policy compliance checking needed to implement these functions.  The point is to show that these trust functions for a federated cloud can be implemented compactly using scripted logical trust  (about 600 lines), and that the resulting implementation is fast enough to use in practice.

In a real deployment, each individual cloud site manager and each control server (e.g., an authority for slices or projects) is a server that possesses an RSA keypair (it is a principal) and runs a private SAFE engine as a local companion process.   Each server uses the REST API to invoke its trust scripts in its local engine through a protected socket.  Each server trusts its local engine and scripts to fetch and validate relevant certificates for its clients, to perform all access checks for its policy, and to generate certificates with its keypair and post them as needed, as programmed in its \system scripts.  

We measure the client-perceived end-to-end latency for canned sequences of operations that implement basic cloud functions as described above. For example, a user U1 creates a project and delegates membership to U2,  who requests to create a slice in the project, and then populates the slice with resources (e.g., VMs), perhaps provisioned from multiple sites and linked together in various ways.    We measure the combined costs for all trust-related functions needed for these sequences: all round-trip script calls, certificate handling, posting/sharing certificates through the shared certificate store, script interpreter costs, and logic query prover/inference costs.   We exclude costs for any actual manipulation of cloud resources (e.g., virtual machine provisioning) that would occur after request authorization is complete: those operations are implemented at a different layer (e.g., ExoGENI/OpenStack), and their costs are independent of the logical trust architecture. 
 
For these experiments we serve the SAFE/\system calls of multiple participating principals on the same engine.  In this way we measure the throughput that each engine can achieve under a heavy logic service mix that is representative of all the trust-related functions for a federated cloud.  In a real deployment these costs are spread across many servers (e.g., one per principal) in parallel: capacity scales with the size of the federation.  The system's only fundamental scaling bottleneck is the underlying certificate store---a scalable key-value store.  However, our bundling approach results in higher ratios in the logic set cache than the cloud servers would see in practice, reducing costs for fetches and signature checking. 

Each SAFE/\system engine instance runs on a four-core KVM (Intel Xeon CPU E5520
@ 2.27GHz) with 12 GB of RAM and 1Gb/s Ethernet.   One-way network delay between two instances is 0.46 ms.   
The certificate store runs on five similar VMs running Riak
2.1.4~\cite{riak} with a replication degree $N=3$ and
$R=1$, $W=3$.    Each posted logic set is materialized as a certificate with a 2048-bit RSA signature.  Tokens and principal IDs are self-certifying 256-bit SHA hashes (44-byte base64-encoded).  The logic payloads of certificates range from 467-840 Unicode characters.  All keypairs are pre-generated.

For these experiments, we created a synthetic federation with a root, ten cloud providers, and two authorities of each type (MA, PA, SA).   We created 20K federation users, 10K of whom are team leaders, 10K projects (one for each leader), 5000 slices from users who have delegated membership in randomly selected projects, and tens of thousands of slivers.  For the stitching experiments we created slices spanning all providers, and linked their dataplanes in rings while creating additional slivers.

Note that the cost for each request depends only on the number of certificates linked into the logic sets that are relevant to it, and the complexity of the \system policy applied to them: the certificate linking abstraction enables a server to identify and fetch the relevant certificates as needed.   In particular, the cost for a request scales with the number of principals involved {\it in that request}---for example, the length of the delegation chain for a slice permission (capability).   The load generator selects principals and objects randomly for each request, so the scale of the system---the total number of principals and objects and the number of participating cloud servers--- influences only the effectiveness of the logic caches in each server, and not the processing costs for each request.   The cost {\it per principal} or {\it per request} of using trust logic is the same at 100K principals/sites or a million principals/sites.



\begin{table}[ht!]
\begin{center}
\begin{footnotesize}
\begin{tabular}{c  c   c  c  c} \hline

\textbf{Cloud operation}   & \textbf{\# \system}          & \textbf{\# \system }              & \textbf{Latency}          & \textbf{Throughput}     \\ 
                                         & \textbf{posts}                 & \textbf{queries}                    & \textbf{(ms)}               & \textbf{(ops/sec)} \\ \hline \hline
\textbf{Create user/PI}     &  4                                    &  N/A                                      & 57.1                               &  104    \\ \hline
\textbf{Lookup user/PI}    & N/A                                 &  1                                          & 7.3/7.8                         &   605/544    \\ \hline
\textbf{Create project}      & 4                                    &   1                                         & 61.4                                 &   107   \\ \hline
\textbf{Delegate project}  & 2                                    &  N/A                                      & 29.1                                    &   206    \\ \hline
\textbf{Lookup project}    & N/A                                 & 1                                           & 7.8                                    &   543    \\ \hline
\textbf{Create slice}         & 4                                     & 1                                          & 71.7                                     &   86       \\ \hline
\textbf{Lookup slice}        & N/A                                 & 1                                          & 7.7                                   & 672      \\ \hline
\textbf{Name slice}          & 2                                     & N/A                                      &  28.3                                  & 209     \\ \hline
\textbf{Create sliver}       & 2                                     &1                                           &  42.8                                   & 178      \\ \hline
\textbf{Intraslice stitch}    & 2                                    & 3                                          &   48.9                                  & 137      \\ \hline
\textbf{Create stitchport} & 4                                     & 1                                          &   73.7                                  & 95      \\ \hline
\textbf{Interslice stitch}    & N/A                                  &2                                           &  21.1                                   & 282     \\ \hline

\end{tabular}
\end{footnotesize}
\end{center}
\caption{\footnotesize{ Latency and throughput of selected high-level cloud operations. We report the 95 percentiles
in each cloud operation scenario.  All measurements are taken from the test harness. Latency includes network delays between the test harness and the logical server and between the logical server and the storage.  Peak throughputs are obtained at concurrency level $C=30$.  Latencies are measured with $C=1$.}}
\label{table:cloud-ops-lat-throughput}
\end{table}

\begin{figure*}[t!]
\centering
\begin{subfigure}{.45\textwidth}
  \centering
    \includegraphics[width=1\textwidth]{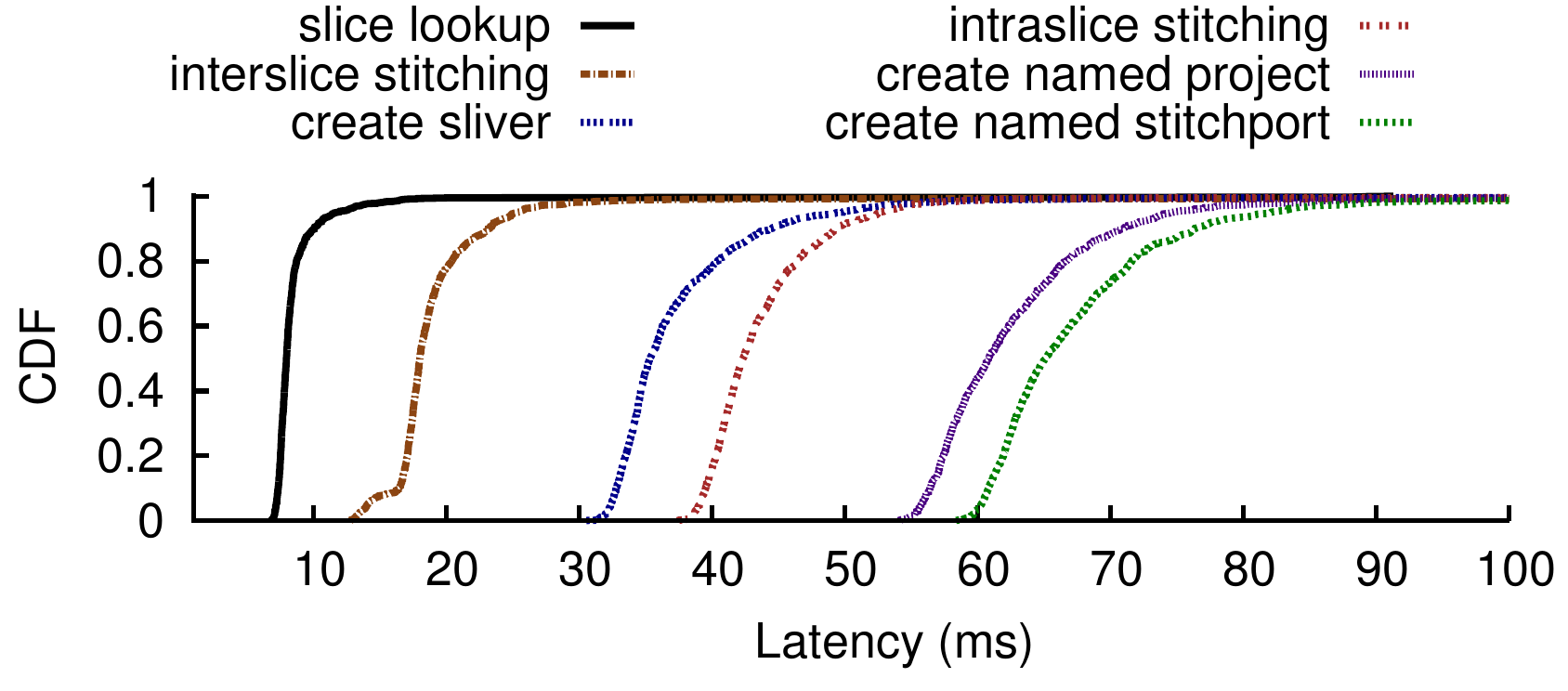}
  \caption{\bf  \footnotesize Latency of a selected set of common
    cloud operations with different costs, implemented as sequences of \system operations. }
  \label{fig:latency-cloud-ops}
\end{subfigure}%
\quad \quad
\begin{subfigure}{.45\textwidth}
  \centering
     \includegraphics[width=1\textwidth]{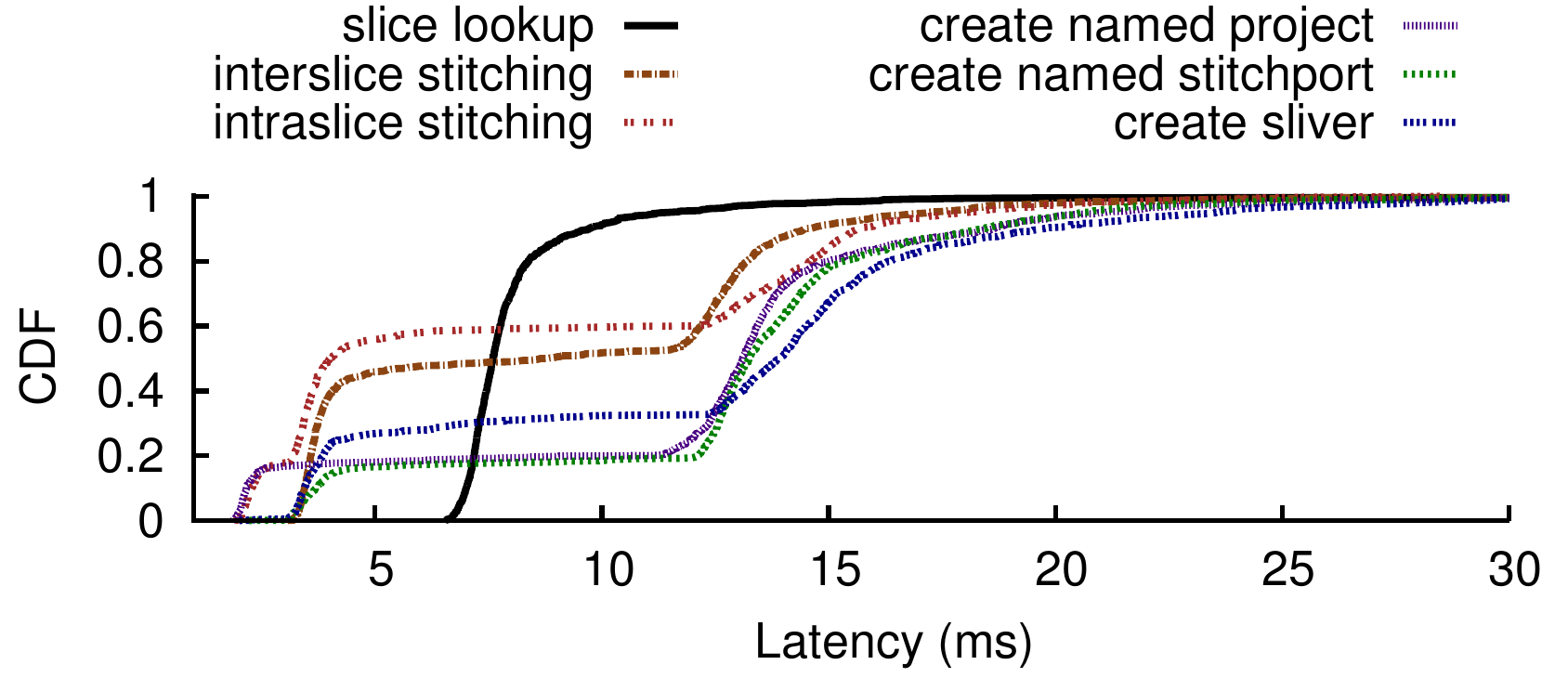}
  \caption{\bf  \footnotesize Latency of \system operations used to implement
    each type of cloud operation, shown as a CDF of per-operation latency for
    the combined \system operation stream.}
  \label{fig:latency-cflo-ops}
\end{subfigure}%
\caption{\bf  \footnotesize Latency of cloud operations and \system operations for synthetic federated trust scenarios.   Underlying \system, operations such as checking compliance and
posting certificates have different cost.  Posting is slower due to signing costs and post latency; compliance checks can be faster due to set/certificate caching.   In \system, end-to-end compliance checking can be completed within 3 ms for most operations. Latency includes network delay and is measured under concurrency level $C=4$.
}
\label{fig:op-latency}
\end{figure*}

Table~\ref{table:cloud-ops-lat-throughput} lists standard operations and their 95\% latencies and peak throughputs on a single 4-core SAFE instance.   Each high-level operation is implemented as a sequence of underlying \system API calls, including those shown in Table~\ref{tab:cflo-api}.  The load generator is multi-threaded, so the sequences are interleaved.  Figure~\ref{fig:op-latency} shows the distribution of latencies at
both granularities: complete sequences, and individual primitive operations within the sequences.
The results reflect latencies in the tens of milliseconds to issue certificates due to signing and posting costs, and much lower costs for the more common verify operations due to caching and other factors, as expected.   Fetch latencies due to cache misses are visible in the latency distributions.

We also performed experiments with multiple SAFE instances in which each principal is assigned randomly to an instance, which performs all operations requested by that principal.   This shows that the code can run in a fully distributed deployment, but the results do not add much insight.  They show additional costs to fetch and share certificates through the shared store; this cost is sublinear in the number of certificates involved in each request, and is determined by access latency to the store.  The fetches are partially parallelized according to the structure of the linked DAG.

These results show that operation costs for logical trust are practical for real deployments.   The logical trust model is flexible and can represent a wide range of trust delegations and access policies concisely.  It makes it possible to build and operate complex federations---and other multi-domain applications---with a small amount of ``extra'' code to capture trust concerns.



\ifpdf

\begin{figure}[t!]
\begin{center}
\epsfig{file=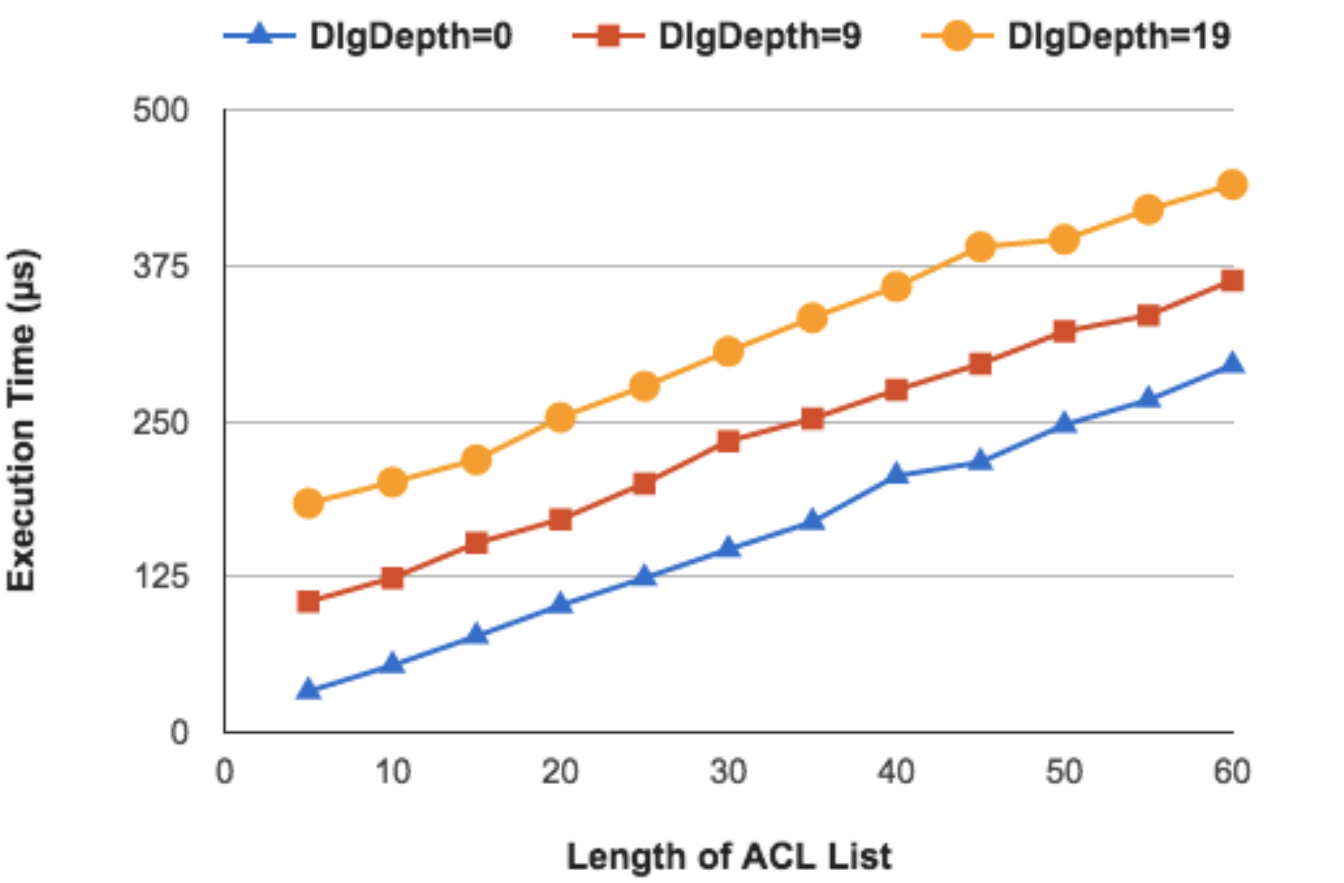, height=5.0cm}
\caption{\bf  \footnotesize  Raw inference time for access checks against an ACL of groups, as a function of ACL length and group delegation depth.   The logical inference costs are a few hundred microseconds for checks that are more complex than are likely to occur in practice. }
\label{fig:safe-time}
\end{center}
\end{figure}

\fi


For these typical operations and scenarios in the cloud federation example, SAFE identifies and retrieves a tightly bounded superset of relevant certificates for each trust decision automatically, and the cost of compliance checks is linear with proof length.  However, more complex policies may show higher costs, particularly for disjunctive policies (complex ACLs, cross-federation with multiple trust anchors).   The multiple branches force the prover to search each branch looking for a proof.  For example, a user request for access may search a long list of groups in an ACL, looking for one that includes the requester.   

To illustrate this concern and focus on the cost of the logical reasoning itself,
Figure~\ref{fig:safe-time} shows the logical inference cost for access
checks  against a list of groups in an ACL, as a function of the length
of the ACL list and the depth of delegation of the
user's membership in a single group in the list.   Costs grow with the number of
disjunctions (ACL length), as well as the cost to traverse the
group delegation chain to form the proof of access.  This delegation cost is linear in SAFE due
to the use of a secondary index.  


Overall, the results suggest that logical inference is cheap in the common case given that the certificate linking structures constructed by the \system scripts focus the prover on relevant logic content, and prune out extraneous statements.
Thus logical trust is cheap in the common case: cost grows with the complexity of the policies, but we pay only for the policy complexity that we use.

%% file: related.tex

{\bf PlanetLab.}  The PlanetLab ~\cite{planetlab:osdi06} network testbed is an early example of a distributed cloud.  The terms {\it slice} and {\it sliver} and our exemplary model of slice-grained access control and signed capability-based delegations for projects and slices---as used in GENI---is derived from PlanetLab.  We show how to implement these (and many other trust features that go beyond PlanetLab) in a unified and flexible logical system that can also capture a wide range of alternatives summarized below.

{\bf Cloud federation standards.} The OGF Open Cloud Computing Interface (OCCI) standard API for cloud services~\cite{ogf-occi}.  An IEEE working group is developing standards for cloud peering (Intercloud Interoperability and Federation IEEE P2302), supported by an Intercloud Testbed Initiative~\cite{intercloud-testbed}.
Papers summarizing the effort and its trust architecture include \cite{bernstein2010intercloud, bernstein2011intercloud}.  Briefly, it proposes federated identity management that encompasses the providers, with common trust anchors (e.g., cloud exchanges) certifying the providers (similar to the trust structure in this paper), and provider groups (trust zones) that reflect varying levels of trust of the providers.  It raises the problem of how to incorporate dynamic trust into certificates issued by the anchors; we show how to solve that problem.


{\bf FIRE.}  The EU-FIRE federation architecture~\cite{vandenberghe2013architecture} plans a similar certification of identity providers and brokering services from a federation trust anchor, and rules-based authorization by participating providers.  BonFIRE~\cite{jofre2014federation} uses a similar structure and supports OCCI.

{\bf Grid.}  The evolution of security architecture for grid computing~\cite{foster01grid} reflects similar concerns and choices.
For example, many deployed grids today bridge 
web single sign-on (SSO) identity services such as Shibboleth~\cite{shib04} to a PKI-based certificate
system for hands-free user control; examples include recent versions of MyProxy~\cite{myproxy}, the Short-Lived Credential Service portal (SLCS), and several others.   GENI MemberAuthority (MA) is similar to these; they are also known as {\it identity brokers}.
Many grid systems employ a service called Virtual Organization Management Service (VOMS~\cite{voms}) to manage user membership in
Virtual Organizations (VOs), which are groupings of principals spanning multiple identity domains.
The VOMS issues credentials as X.509 attribute certificates signed under its own keypair and binding a user's public key to one or more roles scoped to a named VO.   VOs are similar to groups or projects in this paper.

{\bf Logical trust.}  Trust/authorization logic~\cite{abadi93:accesscalculus,lampson92:authentication,spki-rfc2693,jim:sd3:2001,Li02:RT,detreville:binder:2002,Sendlog:Abadi:2007,abadi08, Schneider:2011:NAL:1952982.1952990} is a unifying formalism that can capture these attribute-based mechanisms and policies declaratively and concisely, minimizing the need for custom software, formats, and protocols to implement each design choice.  
The contributions of this paper (and of SAFE)---generalized certificate linking with a common certificate store, programmable scripting, and layered cloud federation---are independent of the trust logic in use.  We prefer to use a standard logic (Datalog) to balance expressive power, tractability, and accessibility for practical use.   In fact, Datalog-with-says is provably the most expressive tractable logic for trust: other logics are either less powerful and lack essential features such as conjunction (SPKI/SDSI~\cite{spki-rfc2693,abac-spki}) or objects (RT0~\cite{Li02:RT}), or are merely syntactic variants of Datalog, or else are intractable and are therefore (in our view) not suited to practical use.  One contribution of this paper is to show that Datalog is sufficient to represent cloud federation needs without these more complex logics.

Grid-inspired research has yielded several PKI-based trust systems that are logical in that they
combine roles and delegations with some
form of declarative policy~\cite{gt4-abac}.  Examples include the PERMIS~\cite{permis02,permis-modular08} system used in European grid initiatives.  These systems generally follow the approach pioneered by SPKI/SDSI, but they introduce custom policy languages.    Most recently, FLANC has been proposed as a custom logic for software-defined network exchanges (SDX)~\cite{Gupta:2016:FLANC}, but it is no more powerful than Datalog with constraints~\cite{Li03:constraint}, or else it is intractable.

{\bf GENI-ABAC.}  GENI uses custom certificate formats and custom validation code to implement its trust model, but alternative support for logical trust exists based on the ABAC software from USC-ISI~\cite{faberauth2016}, which is based on the RT family of logics~\cite{Li02:RT}.  We contributed substantially to the GENI-ABAC design. However, GENI abandoned logical trust in favor of more ad hoc approaches for reasons of expediency in the face of various practical concerns: difficulty in identifying relevant credentials and passing them, difficulty in integrating with established software in multiple languages, and lack of expressiveness.  This paper shows how to address these practical concerns via certificate linking, passing certificates by reference through a shared repository, certificate caching, decoupling of logic concerns from the application into trust scripts, integration of SAFE as a local process that interprets trust scripts and is accessed through a REST API, and use of a Datalog-complete trust logic with a standard syntax and a lightweight service-oriented implementation.

%% file: concl.tex
SAFE is a trust management system that uses a trust logic to represent policies, endorsements, and delegations.   SAFE supports semantically rich certificates and a logic-based authorization engine
implemented in a comprehensive framework that materializes logic sets as certificates and stores them
as linked DAGs in a common key-value store. 
\system uses SAFE to implement the GENI trust and naming model, and extends it to support richer access control, cross-slice peering, and federation peering.


Trust logic is useful as a specification tool for federated cloud architecture, independent of the implementation.  With SAFE, logical trust also enables a practical and concise implementation using declarative policy.   This enables deployments to use a wide range of trust structures and policies specified in declarative logic, using the same software base.  The policies and trust structure may evolve over time without modifying the software.



